\documentclass[pra,notitlepage,showpacs,longbibliography,nofootinbib,superscriptaddress
]{revtex4-2}

\usepackage{amssymb,amsmath,amsthm}
\newtheorem{theorem}{Theorem}%[section]
\theoremstyle{definition}
\newtheorem{definition}{Definition}%[section]

\usepackage{algorithm}
\usepackage{algpseudocode}

\theoremstyle{proposition}
\newtheorem{proposition}{Proposition}%[section]

\theoremstyle{conjecture}
\newtheorem{conjecture}{Conjecture}%[section]

\theoremstyle{remark}

\newtheorem{corollary}{Corollary}[theorem]
\newtheorem{lemma}[theorem]{Lemma}

\usepackage[pdftex]{graphicx}
\usepackage[usenames,dvipsnames]{color}
\usepackage{epstopdf}
\usepackage[hidelinks]{hyperref}
\usepackage{tabularx}

\usepackage{mathrsfs}

\usepackage{soul}

\usepackage{tikz}
\usetikzlibrary{quantikz2}
\usepackage{physics}

\usepackage{amssymb,amsmath,amsthm}

\usepackage{orcidlink}

\newcommand{\p}[1]{\mathbf{p}}
\newcommand{\q}[1]{\mathbf{q}}%momenta

\newcommand{\PP}[1]{\mathrm{(L)}}%particle/anti-particle
\newcommand{\AP}[1]{\mathrm{(R)}}

\newcommand{\w}[1]{\mathbf{\hat w}}

\newcommand{\z}[1]{\theta}
\newcommand{\mix}[1]{\Omega}

\newcommand{\aop}[1]{\hat{a}}
\newcommand{\cop}[1]{\hat{c}}

\newcommand{\sm}[1]{\mathrm{SM}}
\usepackage{slashed}

\usepackage[encapsulated]{CJK}

\usepackage[T1]{fontenc}
\usepackage{beramono}
\usepackage{listings}

\lstdefinelanguage{Julia}%
  {morekeywords={abstract,break,case,catch,const,continue,do,else,elseif,%
      end,export,false,for,function,immutable,import,importall,if,in,%
      macro,module,otherwise,quote,return,switch,true,try,type,typealias,%
      using,while},%
   sensitive=true,%
   alsoother={$},%
   morecomment=[l]\#,%
   morecomment=[n]{\#=}{=\#},%
   morestring=[s]{"}{"},%
   morestring=[m]{'}{'},%
}[keywords,comments,strings]%

\graphicspath{{FIGs/}}
\begin{document}

\title{Quantum computation with the eigenstate thermalization hypothesis instead of wavefunction preparation}
\author{Thomas E.~Baker\orcidlink{0000-0002-3142-0767}}
\email[Please direct correspondence to: ]{bakerte@uvic.ca}
\affiliation{Department of Physics \& Astronomy, University of Victoria, Victoria, British Columbia V8P 5C2, Canada}
\affiliation{Department of Chemistry, University of Victoria, Victoria, British Columbia V8P 5C2, Canada}

\date{\today}

\begin{abstract}

It is proposed that the ability for a quantum circuit to thermalize under time evolution is a valid way to compute linear algebra problems. The algorithm makes use of the eigenstate thermalization hypothesis and full ergodicity in quantum systems to produce an equal superposition of eigenstates. The quantum phase estimation subroutine then allows for the computations of functions of the input operator, leading to a variety of methods in linear algebra. The algorithm circumvents the need for elaborate wavefunction preparation on the quantum computer to find the solution of the linear algebra problem in poly-logarithmic time. 

\end{abstract}

\maketitle

\tableofcontents

\section{Introduction}
If a small container with a gas inside is opened, then our observation in classical statistical physics is that the gas will diffuse into the rest of the room. This process takes an initial configuration that is far from equilibrium and then achieves an equilibrium statistical model. Similarly, pouring milk into a tea will mix the milk uniformly. There is a common feature between these cases. Undoing any of these operations is difficult to the point that the initial state of the system is lost by the end of the problem. The process of reaching the equilibrium value involves in the classical case a lack of retention of the initial state \cite{reif2009fundamentals}. 

Quantum physics, in its simplest form, does not naturally lead to this observation for quantum systems. Since time-dependence is unitary, one should always expect to ``unwind" the time evolution and recover the initial state. This would imply that knowledge of the initial state is always retained by the quantum system.

\subsection{Matching classical and quantum physics}
But if quantum mechanics is the basic description that leads to classical physics, then it is a natural thought to wonder how the quantum system can achieve the ultimate loss of initial state information in the quantum system, while reaching the equilibrium statistical value. Since quantum mechanics ultimately limits to the classical case, then some way of understanding how the quantum system obtains its final answer must be available in nature. How a system achieves equilibrium then must rest on a hypothesis, since the description does not automatically follow from the fundamental postulates of quantum mechanics.

%A full understanding of how quantum systems thermalize is only part of the story. 
%Understanding quantum physics is worthwhile and ever-increasingly more important as methods to control quantum systems. 
\subsection{Computation based on entangled states in quantum physics}
Improved methods of the control of quantum states has opened up the possibility for computation based on quantum properties. The advantage of the quantum computer is that tasks using a full $N$-sized Hilbert space are controlled by $\log_2N=n$ qubits, suggesting that large problems can be exponentially compressed \cite{nielsen2010quantum}. Many quantum algorithms have been proposed, but the bar for deriving new scientific discoveries is high for these algorithms since they have to compete with decades of advancement on classical computers and tested uses for those algorithms. %Still, finding a suitable algorithm on the quantum computer that avoids

%More recently, quantum computers are envisioned to offer a completely new algorithmic paradigm for the computation of large-scale problems. An input operator of size $N\times N$ is represented on $\log_2N=n$ qubits, suggesting that large problems can be exponentially compressed. Quantum computation had many landmark discoveries from the perspective of computer science.

Use of the quantum computer is subject to many constraints, so solving problems that scale polynomially on the classical computer may in principle be expressible on the quantum computer with an exponential advantage. However, certain hidden costs of each step of the quantum algorithm can prevent the algorithm from reaching its theoretical potential.

A vigorous study of which resources are best to use on the quantum computer has been of interest for nearly 30 years. Basing algorithms on effects that we see experimentally in quantum physics is thought to be a viable route to new algorithms. This is the original promise of quantum computing: to use quantum phenomena to compute relevant quantities \cite{nielsen2010quantum}.

%One idea that can be used to make a quantum algorithm for the quantum computer is to use the inherent properties of quantum physics to perform computations that would be difficult on the classical processor. After all, the original goal of quantum computation is to use the fundamental property of entangled states to perform computation \cite{nielsen2010quantum}, so b

%Quantum algorithms

%\begin{theorem}[REMOVE!]
%HERE ME YO!
%\end{theorem}

\subsection{Solving linear algebra with quantum computation}One of the most notable problems to solve on the quantum computer are linear algebra problems of the form
\begin{equation}\label{linearalgebra}
\hat A\mathbf{x}=\mathbf{b}
\end{equation}
for some input vector $\mathbf{b}$ and desired output $\mathbf{x}$ applies to a wide number of problems. The ability for a quantum computer to compute this type of problem is a natural extension of the tested application of linear algebra techniques to describe quantum systems \cite{townsend2000modern}. So, using a quantum computer to solving linear algebra is highly sought.

On the classical computer, the solution can be obtained by taking an LU decomposition and use the characteristic form of the upper- and lower-triangular matrices, which costs $O(N^3)$ time in general, to find the inverse $\hat A^{-1}$ to give $\mathbf{x}=\hat A^{-1}\mathbf{b}$ \cite{press1992numerical}. 

The quantum computer has been conjectured to run the problem in poly-logarithmic time \cite{harrow2009quantum}, but several steps require extra overhead for the most general problem. Of principle importance here is the need to make an initial wavefunction on the quantum computer. This can be expensive to create and reduce the viability of the algorithm depending on the use case.

In general, the fine details of the quantum algorithms sometimes reveal parts where the algorithm is secretly inefficient. If a procedure is used that requires $O(N)$ steps is needed in an algorithm (such as a full tomography) then an algorithm does not achieve a formal poly-logarithmic scaling in practice.

The most glaring issue that is proposed to be removed here is that of wavefunction preparation on the quantum processor \cite{aaronson2015read}. Wavefunction preparation should generally cost $O(N)$ operations, although some cases can be found that tend more towards $O(n)$ which is far more efficient. The question of interest here is to wonder if no wavefunction preparation was required for any case.

\subsection{Combining the above: thermalizing quantum circuits}
A dual algorithm that is not reliant on wavefunction preparation for solving linear algebra problem is proposed here. Instead of using wavefunction preparation, we use the thermalization of an arbitrary, non-eigenstate to--on average--generate an equal superposition of eigenstates, which can be used to compute the eventual expectation values. A key ingredient that is required to ensure that all states are equally probable. Throughout, this will be referred to as fully ergodicity--where all states in the Hilbert space are equally likely--and is what the state will appear as on a time-average.\footnote{Note a difference in the definition here, the ergodicity here is a full ergodicity over the entire Hilbert space. Ergodicity in other works refers to equal probability over a single symmetry sector.}

The idea in this paper is to combine a) the thermalization of quantum systems to an ergodic state and b) a quantum phase estimation to weight the eventual measurement, giving access to a function of an input operator, including the inverse. 
%-to generate an equal superposition of eigenstates on the order of the thermalization time--and then use
This will produce an expectation value of an input operator while erasing the information of the input state. Effectively, this allows for a good approximation of the trace over an equal superposition of eigenstates. The strategy generalizes to many cases as will be shown.

%costs and tradeoffs

The concepts hinted at in this introduction are formalized with mathematics with an overview of known key results and their motivation in Sec.~\ref{motivation} with background knowledge in Sec.~\ref{motivation}. A review of the eigenstate thermalization hypothesis (ETH) is then given in Sec.~\ref{ETHreview}, followed by its application to circuits in Sec.~\ref{ETHcircuits}. Some examples for linear algebra quantities are given in Sec.~\ref{examples}. The last section, Sec.~\ref{QPEissue} discusses systematic errors in this algorithm from the quantum phase estimation and potential solutions.

%The remarkable fact of the quantum computer is that the input wavefunction contains up to $N$ values but it is only represented on $n=\log_2N$ qubits. Therefore, the opportunity to solve problems with a 

%The fine details of the quantum algorithms sometimes reveal parts where the algorithm is secretly inefficient. If a procedure is used that requires $O(N)$ steps (such as a full tomography) then an algorithm does not achieve a formal poly-logarithmic scaling in practice.

%\section{Summary of results}
%
%The main outcome from this paper is to demonstrate that 

\section{Motivation}\label{motivation}

In general, the form that a measurement in quantum mechanics will take is 
\begin{equation}\label{Oexpect}
\langle\hat{\mathcal{O}}\rangle=\mathrm{Tr}\left(\hat\rho\hat{\mathcal{O}}\right)
\end{equation}
where $\hat{\mathcal{O}}$ is some operator and $\hat\rho$ is the density matrix. The trace of the product of the two operators gives the expectation value. 

The core development in this paper is to propose that an equal superposition of eigenstates can be created as an ensemble (time-average) in $O(n)$ time for a wide-class of models through thermalization. Combining thermalization with the quantum phase estimation (QPE) for an input $\hat A$ allows for the expectation value of a function of the input matrix, $\hat{\mathcal{O}}=f(\hat{A})$, by reweighting eigenvalues ({\it i.e.}, $\hat{\mathcal{O}}=\hat A^{-1}$). The key focus of this work is to generate an algorithm for a matrix $\hat A$ representing a linear algebra system, which often lacks the structure of a physics problem and can be taken to be non-integrable.

The wavefunction-based preparation methodology is discussed in a way that leads to an understanding of the dual formulation of those methods.

\subsection{Wavefunction based methods}
For methods of linear algebra that rely on wavefunction preparation, one supposes that the first step on the quantum computer is to construct the starting wavefunction, $\Psi$. In terms of Eq.~\eqref{Oexpect}, the appropriate form would have $\hat\rho=|\Psi\rangle\langle\Psi|$ and therefore $\langle\hat{\mathcal{O}}\rangle=\langle\Psi|\hat{\mathcal{O}}|\Psi\rangle$. 

Often in quantum computation, a wavefunction must be prepared on the quantum computer. This is expected to be exponentially costly in the most general case \cite{aaronson2015read}.  Using such an operation for a quantum algorithm would therefore cost $O(N)$ steps \cite{aaronson2015read}, which is not the poly-logarithmic scaling that is desired for these problems. If general wavefunction preparation is required, then an efficient algorithm can only happen on a restricted set of cases \cite{tang2021quantum} since the wavefunction formulation will be strongly dependent on the cost to construct a wavefunction.\footnote{Albeit, those specific instances may cover quantum problems where assumptions such as locality may be assumed to constrain what types of quantum problems can be created. Even in this case, removing the wavefunction preparation step would be beneficial and it will be seen in this construction that all possible input wavefunctions can be solved in a single instance of this algorithm.}

\subsection{Dual formulation}
The dual formulation would be to define a set of eigenstates of another operator $\hat{A}$ as
\begin{equation}\label{operatorA}
\hat A=\sum_kE_k|k\rangle\langle k|
\end{equation}
in terms of eigenvalues $E_k$ and eigenvectors of $\hat{A}$, $|k\rangle$, such that $\hat{A}|k\rangle=E_k|k\rangle$. The preoccupation here is when 
\begin{equation}
\hat{\mathcal{O}}=f\left(\hat A\right)=\sum_{k=1}^Nf(E_k)|k\rangle\langle k|
\end{equation}
or that $\hat{\mathcal{O}}$ is a function $f$ of $\hat A$. For example, $\hat{\mathcal{O}}=\hat A^{-1}$ would imply that $f(E_k)=1/E_k$. This form broadly applies to many examples as will be shown.

The expectation value from Eq.~\eqref{Oexpect} now takes the form of 
\begin{equation}\label{dualform}
\langle\hat{\mathcal{O}}\rangle=\sum_{k=1}^Nf(E_k)\langle k|\hat\rho |k\rangle
\end{equation}
where now all possible eigenstates must be summed in order to generate all possible expectation values $\langle k|\hat\rho|k\rangle$ in a single expectation value. %The wavefunction that was prepared in the direct formulation for wavefunction preparation above is now contained as an operator in $\hat\rho$. 

\begin{figure}
	\begin{algorithm}[H]
		\caption{Overview of ETH-$\Sigma$}
		\label{ETHsum_overview}
		\begin{algorithmic}[1]
			\State Generate an equal superposition of eigenstates, Eq.~\eqref{equalsuperposition} \Comment{By thermalization, Conjecture~\eqref{ETH}}
		\State Apply a quantum phase estimation \Comment{As in Def.~\eqref{QPE}}
		\State Measure $\check{\mathcal{O}}=\hat\Delta\otimes\hat\Upsilon$\Comment{$\hat\Delta=\hat\rho$ and $\hat\Upsilon$ contains $f(E_k)$ as in Eq.~\eqref{dualform}}
	\end{algorithmic}
\end{algorithm}
\end{figure}

\section{Main result: summation of all states by the eigenstate thermalization hypothesis}

The primary proposal of this paper is summarized as the following proposition.\footnote{ETH itself is not formally proven, hence basing the main result on ETH should properly on the same footing as ETH. Should ETH ever be proven, the proposition would follow by construction in this paper and apply when ETH applies.} An overbar is used to denote a time-average.

%The main results is summarized. 

\begin{proposition}[Thermalization of a the quantum circuit]\label{mainproposition}

A time evolution of an arbitrary, non-eigenstate $|\Psi(t)\rangle=e^{i\hat At}|r\rangle$ in a quantum circuit will approximate an equal superposition of eigenstates in a time-averaged ensemble of states. The resulting micro-canonical ensemble of an operator $\hat\rho\hat{\mathcal{O}}$ will be
\begin{equation}\label{conjecture1}
\overline{\langle \Psi(t)|\hat\rho\hat{\mathcal{O}}|\Psi(t)\rangle}\approx\frac1N\mathrm{Tr}\left(\hat\Gamma\hat\rho\hat{\mathcal{O}}\right)=\frac1N\sum_{k=1}^Nf(E_k)\langle k|\hat\rho|k\rangle\Gamma_{kk}
\end{equation}
for a density matrix $\hat\rho$ containing the initial wavefunction for a problem as an operator, an ensemble density matrix $\hat\Gamma$ which for infinite precision is the identity but restricted at finite precision. %This is equivalent to thermalizing to an equal superposition of eigenstates, $|\Xi\rangle$.
\end{proposition}

The implication of $\Gamma_{kk}$ in Eq.~\eqref{conjecture1} is that it defines a range states which have values above the minimum precision requested in the quantum computation. Around a Gaussian distribution, as is typical in statistical physics systems, this could reduce the sum over all $N$ states to a window over only $\mathcal N$ states,
\begin{equation}
\frac1N\mathrm{Tr}\left(\hat\Gamma\hat\rho\hat{\mathcal{O}}\right)=\frac1N\sum_{k=1}^Nf(E_k)\langle k|\hat\rho|k\rangle\Gamma_{kk}\approx\frac1{\mathcal N}\sum_{f(E_k)\langle k|\hat\rho|k\rangle>\varepsilon}f(E_k)\langle k|\hat\rho|k\rangle
\end{equation}
and recover the micro-canonical ensemble, $\mathcal{N}$ representing the number of states kept to the level of precision requested. An approximation appears because of how sharply the cutoff between states above the precision limit and below it are represented, effectively cutting off digits of the summed values for higher precisions $1/\varepsilon$.

In order to compute Eq.~\eqref{dualform} based on this proposition, the steps in Algorithm~\ref{ETHsum_overview} are proposed. This algorithm is known as ETH-$\Sigma$ (``ETH-summation") and is derived in the following in two forms after reviewing ETH and applying it to quantum circuits. %The general algorithm defines a generic operator $\hat\Delta$ which is equal to $\hat\rho$ in Eq.~\eqref{dualform}. The elements of $\hat\Upsilon$ contain $f(E_k)$ which will be elaborated on.

The overall scaling will be $O(\frac\tau{\varepsilon^\gamma}\log_2N)$ for a thermalization time $\tau$, number of qubits $n=\log_2N$ when the time evolution operator is efficiently implementable, and $\gamma=1$ or 2. Evidence for ETH indicates that the thermalization time will happen in poly-logarithmic time, although there are exceptions ({\it e.g.}, many-body localization which has an infinite thermalization time).

The basic elements necessary to compute Eq.~\eqref{dualform} on the quantum computer are developed in the next sections.

This concludes the review of the background material for the ETH-$\Sigma$ (``ETH-summation") algorithm.

\section{Background concepts}\label{overview}

This section will cover background information and introduce definitions and foundational results that are necessary for understanding the primary results later on. The content covers statistical physics, quantum physics, and quantum computation.

%\subsection{Core idea: Generation of an equal superposition of eigenstates by thermalization}
\subsection{Time evolutions and thermalization}

Starting with an arbitrary, non-eigenstate wavefunction, the state can be expressed in a superposition of eigenstates as
\begin{equation}\label{arbwavefunction}
|r\rangle=\sum_{k=1}^Nc_k|k\rangle
\end{equation}
where $c_k\neq1$ for any $k$ assuming proper normalization. The initial state dependence contained in the coefficients $c_k$ must be eliminated to serve as a useful computational resource. %The time evolution of $|r\rangle$ is given by
%\begin{equation}
%|\Psi(t)\rangle=\exp(-i\hat At)|r\rangle
%\end{equation}
%which uses the operator $\hat A$ introduced originally.

The proposal here is to use a time-average where the time-evolution is defined as
\begin{equation}\label{timeEvolR}
|\Psi(t)\rangle=\exp(-i\hat At)|r\rangle
\end{equation}
with respect to the operator $\hat A$. It is through a time-average of an operator with respect to $\Psi(t)$ that a sufficient quantum resource can be derived.

%which is the same form as Eq.~\eqref{superposition} but this is meant to be an arbitrary non-eigenstate, not necessarily an eigenstate as in Eq.~\eqref{superposition}.

By itself, Eq.~\eqref{Oexpect} relates to a single state contained in the density matrix $\hat\rho$. There is the concept of an ensemble which is defined as follows.

\begin{definition}[Ensemble]\label{ensembledef}
An average over many states.
\end{definition}

It is over an ensemble of time-averaged states that a system will be allowed to free itself of the initial state dependence.

The definition of the {\it ensemble average} is given by $\frac1{\mathcal{N}}\mathrm{Tr}\left(\hat\Gamma\hat{\mathcal{O}}\right)$ which looks functionally identical to Eq.~\eqref{Oexpect} but there is a qualitative difference denoted by the replacement of $\hat\rho$ (single state density matrix) to $\hat\Gamma$ (ensemble density matrix). The $\hat\Gamma$ operator denotes an average over many states and is not normalized while $\hat\rho$ is normalized, hence there is an additional factor of $1/\mathcal{N}$ where $\mathcal{N}$ is the number of non-zero states in $\hat\Gamma$.

\begin{definition}[Thermalization]
The process by which a system reaches the ensemble average in equilibrium.
\end{definition} 

One can naturally expect that the initial state of a system (perhaps milk pouring into tea) has some ensemble expectation value that is not the equilibrium result.  Thermalization is the process of mixing the milk into the tea, where the information about the initial state of the milk should be lost. In general, a quantum system will thermalize to the micro-canonical ensemble which is defined as follows. This is the elimination of the initial state dependence that is found useful here.
%There is a proposal for how quantum systems thermalize given in terms of the micro-canonical ensemble average.

The formal result that a system will obtain during thermalization is given by a micro-canonical ensemble.

\begin{definition}[Micro-canonical ensemble]\label{microcanonicalensemble}

A micro-canonical ensemble can be written as \cite{reif2009fundamentals}
\begin{equation}\label{MCdef}
\langle\hat{\mathcal{O}}\rangle_\mathrm{MC}=\frac1{N}\mathrm{Tr}\left(\hat\Gamma\hat{\mathcal{O}}\right)=\frac1{\mathcal{N}}\sum_{E<E_k<E+\Delta E}\langle k|\hat{\mathcal{O}}|k\rangle
\end{equation}
where $\Delta E$ is the width of an energy window starting at some energy $E$, ${\mathcal{N}}$ is the number of states in that window, and
\begin{equation}\label{gammadef}
\hat\Gamma=\begin{cases}
1 & E<E_k<E+\Delta E\\
0 & \mathrm{otherwise}
\end{cases}
\end{equation}
is the density operator in the micro-canonical ensemble. The trace in Eq.~\eqref{MCdef} goes over all $N$ ($\ggg{\mathcal{N}}$) states in the Hilbert space.

\end{definition}

The energy window of width $\Delta E$ is often adjusted in numerical studies to make the ensemble-averaged value match the average of the Gaussian distribution.

\subsection{Eigenstate thermalization and random matrix theory}

The statement of how a quantum system thermalizes is given by the eigenstate thermalization hypothesis (ETH).\footnote{In order to formalize the concepts for the quantum information community and to aid adoption of this algorithm, statements from the broader ETH literature are formalized into math headings. There are a variety of other ways to formulate the question into these headers.}% For example, many will primarily attribute ETH to be what is called here Lemma~\ref{RMTconj}, but the core assertion taken here for ETH is that the statement of what the time average goes to the micro-canonical ensemble.}

\begin{conjecture}[Eigenstate thermalization hypothesis]\label{ETH}
The eigenstate thermalization hypothesis supposes that quantum systems thermalize to a micro-canonical ensemble average
\begin{equation}
\overline{\langle\Psi(t)|\hat{\mathcal{O}}|\Psi(t)\rangle}=\langle\hat{\mathcal{O}}\rangle_\mathrm{MC}
\end{equation}
as a general property of some quantum systems.
\end{conjecture}

Interpreted in another way, ETH establishes how the initial state dependence of a system is removed in the ensemble of time-averaged states, although it may not be obvious from the form in Conjecture~\ref{ETH} without more development of the terms behind the statement. Various explanations have been proposed \cite{deutsch1991quantum,deutsch2018eigenstate} and some cases can be solved explicitly to show the relationship under assumptions that allow for an analysis with random matrix theory \cite{srednicki1994chaos,d2016quantum}. Random matrix theory is based on the notion that energy levels can be randomly spaced in a quantum system. % \textcolor{red}{continue here after reading more about random matrix theory}

RMT is a tool to study quantum chaos, the recovery of classical chaos \cite{goldstein2014classical} in a quantum system which is largely an open question. Why should a matrix with randomly spaced eigenvalues give rise to thermalization and quantum chaos? One answer is that the energy-time uncertainty principle $\Delta E\Delta t\geq\hbar/2$ should apply for an observation of any system and that observation should take a finite amount of time, $\Delta t$. The energy width $\Delta E$ observed in a time window would have trouble distinguishing between energies as a fundamental fact, especially if they are closely spaced or randomly spaced. Attempting to trace back a signal to individual eigenstates would require a long $\Delta t$ to get a smaller $\Delta E$ value. Thus, the system can scramble the information amongst similar states on observation. Hence, the usefulness of supposing that the energy levels are randomly spaced, which is effectively the useful approximation to make.

The core assertion (ETH) that the thermalization of quantum states rests upon is not proven definitively for general systems. It does have 30 years of evidence to support the thermalization of quantum states, including real experimental evidence where isolated quantum systems thermalize and when they do not \cite{d2016quantum}. Because of the lack of a proof for the hypothesis, this paper will focus on conjecture. The conjecture is that this is the behaviour of quantum systems in general, although the specific proof strategy used here will be the commonly accepted method of using random matrix theory.

Theorems will not appear since this is a conjecture that is made. Lemmas will be used when a theorem is proven in another paper and corollaries are direct consequences of lemmas. Because the justification for this statement is formalized on the ETH (which is formally a conjecture at this point), the main statement here will be given as a proposition based on ETH as justified through random matrix theory. 

\subsection{Ensemble equivalence}

%We find it useful to discuss the value that a system will thermalize to in terms of the end-density matrix that is required in the eventual time average of the system. In Def.~\ref{microcanonicalensemble}, the primary quantity computed in systems that display ETH, the wavefunction that would need to be prepared to equal the time average would need to give the correct form of Eq.~\eqref{gammadef}. Such an initial state when applied to find the expectation value of an operator $\hat{\mathcal{O}}$ would give the same value as the long-time average.

Equivalently, the micro-canonical ensemble is not the only way to describe a system. The canonical ensemble \cite{reif2009fundamentals} is an equivalent way to describe a thermodynamic system. In this case, the states are weighted by a thermal probability that is dependent on a value $\beta$ (inverse temperature). This formally assigns a temperature to a given system's fluctuations.% that is often fit to the energy statements in the problem.

\begin{definition}[Canonical ensemble]
A canonical ensemble has an inverse temperature $\beta$
\begin{equation}
\frac1Z\mathrm{Tr}\left(\hat\Gamma\hat{\mathcal{O}}\right)=\frac1Z\sum_{k=1}^Ne^{-\beta E_k}\mathcal{O}_{kk}
\end{equation}
where
\begin{equation}\label{boltzmannprobability}
\hat\Gamma=e^{-\beta\hat A}%\mathcal{H}}
\end{equation}
which contains all of the probabilities for the thermodynamical system. The quantity $Z$ is defined as
\begin{equation}
Z=\sum_{k=1}^N\exp(-\beta E_k)=\mathrm{Tr}\left(\hat\Gamma\right)=\mathrm{Tr}\left(e^{-\beta\hat A}\right)
\end{equation}
which is the {\it partition function}.
\end{definition}

The choice of an ensemble in thermodynamics should not alter the final result. Thus, a computation with a micro-canonical ensemble average should have a representation with a canonical ensemble. The equivalence between ensembles means an effective temperature for the micro-canonical ensemble can be assigned for Conjecture~\ref{ETH}.\footnote{In numerical studies the value of $\beta$ is fit to the fluctuations of the time average.}

\subsection{Fully ergodic thermalization}% in quantum circuits}

%Encapsulated in Eq.~\eqref{dualform}, the most useful resource for quantum computation is when the multi-canonical ensemble has an equal superposition of eigenstates of the input operator.

Based on the arguments around Eq.~\eqref{dualform}, an equal superposition of eigenstates would be the most useful state to thermalize to. This is one statement stronger than ETH but one that can be ultimately supplied by ETH and consistent with a multi-canonical ensemble if precision is taken into account as will be shown. %Not only is the system thermalized but it is also ergodic. 

An equal superposition of eigenstates is that 
%Often the most useful condition that is required for general computation--see Eq.~\eqref{arbwavefunction}--that 
each of $c_k\rightarrow1/\sqrt{N}$ in Eq.~\eqref{arbwavefunction} which implies that all states of the phase space are equally probable.\footnote{Some would refer to this case as quantum chaos, but this term will be avoided due to a lack of a firm definition in the literature for quantum systems. The case sought is definitely ergodic since all states are equally probable.}

%Our preoccupation in this work is with a special case of Eq.~\eqref{superposition} where all weights are equal.

\begin{definition}[Equal superposition of eigenstates]
An input superposition of eigenstates that are of equal probability
\begin{equation}\label{equalsuperposition}
|\Xi\rangle=\frac1{\sqrt N}\sum_{k=1}^N|k\rangle
\end{equation}
which has equal superposition in all states and can be considered fully ergodic.\footnote{The term ergodic is used here to indicate equal probability over the entire Hilbert space, not just along a restricted shell as in Liouville's theorem. Another definition of the term appears in some other literature \cite{deutsch1991quantum}.}
\end{definition}

In one sense, if there was easy access to a quantum resource of the form of Eq.~\eqref{equalsuperposition}, $|\Xi\rangle$, then there would be no need for thermalization. One would simply measure $\langle\Xi|\hat{\mathcal{O}}|\Xi\rangle$.  Lacking a readily available wavefunction $|\Xi\rangle$ that can be created on the quantum computer in $O(n)$, the proposal here is to use thermalization to drive the system such that $\beta=0$ (infinite temperature). 

\subsection{Difference in physics between generating a single state and an ensemble}
In the time average, the system would thermalize to the equal superposition of eigenstates, now expressed as an ensemble of states. There is a significant difference in the sense of physics between taking the expectation values as
\begin{equation}\label{state_vs_ensemble}
\langle\Xi|\hat{\mathcal{O}}|\Xi\rangle\quad\mathrm{or}\quad\mathrm{Tr}\left(\hat\rho\hat{\mathcal{O}}\right)\quad\left(\mathrm{with}\quad\hat\rho=\frac1{N}\sum_k|k\rangle\langle k|\right)
\end{equation}
where the density matrix for the first option is a pure state and the density matrix for the second option is a mixed state. The important distinction between thinking of the expectation value as a single state $|\Xi\rangle$ has a different entropy ($S=0$) from the ensemble expectation value ($S=\ln N$, the maximal amount) and so it is not proper to mix the two descriptions. However, since the field is currently investigating wavefunction methods where something like $|\Xi\rangle$ is the starting point for an eventual measurement, the language may percolate into the new discussion here about using time-averaged ensembles.

Because of the traditional focus on wavefunction preparation, saying that one would ``thermalize to a state" should be understood to mean that the system takes--on average--an ensemble that represents an equal superposition of states. It is just a shift in thinking between wavefunctions by themselves and ensembles. This phrase would imply that if the correct state is prepared and used to measure the operator, then the expectation value will be equivalent to the time-average.

%The equivalent wavefunction for this ensemble would give a density matrix of the form $\hat\rho=|\Xi\rangle\langle\Xi|$. %Occaisionally, the time average may be referred to as ``thermalizing to a state" since the time-average is equivalent to a carefully prepared state. This would imply that if the correct state is prepared and used to measure the operator, then this expectation value will be equivalent to the long-time average.
%
\subsection{Choice of operator for fully ergodic thermalization}

How can an arbitrary system be thermalized to an equal superposition of eigenstates? Thermalizing to an equal superposition of eigenstates would be equivalent to Eq.~\eqref{state_vs_ensemble}. The process of thermalization required here is to turn $|r\rangle\rightarrow|\Xi\rangle$ (or more properly as a density matrix $\hat\rho$) in a time-average for a specific choice of $\hat A$. 

%So far, all time evolutions have been framed in terms of a time evolution based on the operator $\hat A$. 
To achieve the final state $|\Xi\rangle$, a specific class of operators $\hat A$ could be chosen or modified to be kicked harder. This concept is natural to the area of open-dynamics. When running a simulation if a chaotic state is not achieved for some purpose, then a stronger ``kick" is introduced to induce quantum chaos. For here, a ``kick" is defined in the sense of a kicked-top Ising model \cite{mumford2025characterizing}. 

Where an operator $\hat A$ possesses symmetries, the time-evolution can be modified to break those symmetries and introduce full ergodicity over the entire Hilbert space. This will provide a time evolution that ``kicks" the system sufficiently hard to have equal probability of being found in all eigenstates during the time-evolution. For example, when the system does not naturally give full ergodicity ({\it i.e.}, when $c_k$ values are not all equal), one can induce this condition by adding kicks to the input operator and then turning it off so that the final answer is representative of a function of $\hat A$ instead of the operator plus some perturbation. The main goal here is to invert the matrix, not necessarily to leave in place the symmetries of the system at all times for study of the physical system. The actual protocol for each system will depend on future use applications.\footnote{Near-term quantum computers may have an advantage due to noise that may induce more transitions between symmetry sectors.}

Introducing sufficient kick terms or additional elements can cause the state to thermalize fully ergodically across all states in the Hilbert space. %The modification of the time evolution will not affect the following, and it is likely that the form of this will need to be determined on a case by case basis. %One way to frame the thermalization of the state is to remove the initial state dependence. The leading framework that--when satisfied--is sufficient to describe thermalization is the ETH. This will be defined in the following sections after describing the broad outlines of the algorithm framework in this section.
The form of Eq.~\eqref{equalsuperposition} is proposed in this paper to be $O(n)$ for a broad class of problems. For some systems, which we will call many-body localized (MBL) \cite{nandkishore2015many,sels2021dynamical}, the time to thermalize is too long to ever reasonably see it thermalize. For many cases, MBL is not present and the algorithm often thermalizes in less than $O(N)$ operations.% \cite{jeffrey}.

%Less is known about ergodicity in quantum mechanics outside of the quantum ergodic hypothesis \cite{}, which applies for semiclassical systems. The ability for the quantum computer to initialize many trajectories at once is a reasonable starting point to allow for ergodicity. We touch on the case of symmetries and quantum scars in the realization of these ideas in a quantum algorithm. The dual algorithm formulated here will require a specific resource as an equal superposition of eigenstates. 

\subsection{Computations on functions of an input operator}

Having established that an equal superposition of eigenstates can be obtained from thermalization, one additional ingredient can be added to obtain the trace over functions of operators. The addition of the quantum phase estimation (QPE) \cite{nielsen2010quantum} will provide a means to weight eigenvalues and utilize the equal superposition of eigenstates to generally solve linear algebra problems. 

%Having established the ideas of thermalization--which would only require time evolution and measurement on the quantum computer--and its connection to linear algebra problems, an additional subroutine can be added to further generalize the algorithm in terms of the quantum phase estimation (QPE).

%The operator $\hat A$ is represented here. However, there is a straightforward way to represent the inverse (or any function of the input matrix $f(\hat A)$) through the quantum phase estimation (QPE) on the quantum computer.

\begin{definition}[Quantum phase estimation]\label{QPE}
The quantum phase estimation takes an input eigenvector $|k\rangle$ and returns a eigenvalue $E_k$ represented on $m$ qubits of precision which we denote as
\begin{equation}\label{QPEaction}
|k\rangle|0\rangle^{\otimes m}\overset{\mathrm{QPE}}\longrightarrow|k\rangle|E_k\rangle
\end{equation}
where $\hat A|k\rangle=E_k|k\rangle$.
\end{definition}

\begin{figure}
\begin{center}
\begin{quantikz}[column sep=14, wire types = {q, q, q, q}]
\lstick{$\ket{0}$}&\gate{\hat H} & & & \phase{} & \gate[3]{\mathrm{QFT}^\dagger}& \\
\lstick{$\ket{0}$}&\gate{\hat H}& & \phase{}  & \wire[u][1]{q} &&\\
\lstick{$\ket{0}$}&\gate{\hat H}& \phase{} & \wire[u][1]{q} & \wire[u][1]{q} &&\\
\lstick{$\ket{\Psi}$}&\qwbundle[Strike Height= 0.2]{n}& \gate{\hat U^{2^0}} \wire[u][1]{q} & \gate{\hat U^{2^1}} \wire[u][1]{q} & \gate{\hat U^{2^2}} \wire[u][1]{q} & &
\end{quantikz}
\end{center}
\caption{The circuit for the quantum phase estimation on $m=3$ qubits. The gates and result are defined in the text with $\hat U=\exp(i\hat A)$. This creates the first three digits of a bit-string representation of the eigenvalue related to each eigenstate in a spectral decomposition of the input state $|\Psi\rangle$.
\label{QPEcircuit}
}
\end{figure}

The QPE applies to any eigenstate or linear combination of eigenstates so long as the operator $\exp(i\hat A)$ is available. Figure~\ref{QPEcircuit} shows the quantum circuit that corresponds to the QPE.

Through the use of the QPE, there is now an auxiliary register containing bit-strings corresponding to eigenvalues of the input operator. These bit-strings can be acted upon by an operator $\hat\Upsilon$. The output of the QPE is to provide weights of bit-strings in the register $|E_k\rangle$ above for the corresponding eigenvalue $E_k$. The operator $\hat\Upsilon$ effectively provides a metric for the bit-strings.% which can be taken to be the energy corresponding bit-string, but it can be generalized.% chosen to be the energy.

But the elements of an operator $\hat \Upsilon$ can be extended to a function of the energies as well.
\begin{definition}[Eigenvalue weighting matrix]\label{UpsilonDefinition}
A matrix $\hat\Upsilon$ will have elements $f(E_k)$ (which can be the identity, $f(E_k)=E_k$, or any other function, including those not strictly dependent on the eigenvalue) for some function $f$ of the eigenvalues $E_k$.
\begin{equation}\label{Upsilon_operation}
\hat\Upsilon=\mathrm{diag}\Big(f(E_1),f(E_2),\ldots,f(E_M)\Big)=\left(\begin{array}{cccc}
f(E_1) & 0 & \ldots & 0\\
0 & f(E_2) & \ddots & 0\\
\vdots & \ddots &\ddots & 0\\
0 & 0 &0 & f(E_M)\\
\end{array}\right)
\end{equation}
where $M\equiv2^m$.
\end{definition}

%However, in general, the expectation value does not necessarily have to have $E_k$ be the factor outside of the expectation value in each term. The term $E_k$ can be promoted to $f(E_k)$ in the elements of $\hat\Upsilon$ to given an expectation value of a function of $\hat A$, $f(\hat A)$. 

The quantum computer performs this operation in superposition, so the QPE can be applied on a time-averaged ensemble that averages to an equal superposition of eigenstates\footnote{From the wavefunction perspective, an abuse of language might be tolerated to say that the time-average resembles the generation of $|\Xi\rangle$ on average. This is not formally correct in terms of statistical physics, but it is natural from the quantum computing perspective. See the discussion around Eq.~\eqref{state_vs_ensemble}.} after thermalization and obtain all of the weights of the eigenvalues in one operation. When taking a function of the eigenvalues, an operator $\hat\Upsilon$ can be generated to measure on the auxiliary register for the eigenvalues $E_k$. The $\hat\Upsilon$ operator applies on an auxiliary precision register of $m$ qubits, which is smaller than the system register of $n$ qubits for most large applications of interest. This implies that a measurement over $\hat\Upsilon$ is contained to only a few qubits ($N\ggg M$).

%\subsection{Relation to the microcanonical ensemble}
%
%What Eq.~\eqref{dualform} does not show, but forms a very subtle point in the construction here is the role of numerical precision. Values of $f(E_k)\langle k|\hat\rho |k\rangle$ is too low in comparison to the precision $\varepsilon$ and so we can include numerical precision as
%\begin{equation}\label{Omc}
%\langle\hat{\mathcal{O}}\rangle_\mathrm{MC}=\mathrm{Tr}\left(\hat\Gamma\hat\rho\hat{\mathcal{O}}\right)=\sum_{k=1}^Nf(E_k)\langle k|\hat\Gamma\hat\rho |k\rangle
%\end{equation}
%where $\hat\Gamma$ is a density matrix  for the microcanonical ensemble.
%%\begin{equation}
%%\hat\Gamma=\begin{cases}
%%1 & E<E_k<E+\Delta E\\
%%0 & \mathrm{otherwise}
%%\end{cases}
%%\end{equation}
%%which connects the expectation value to the micro-canonical ensemble (ME), the true prediction by ETH. This is only a proposed relationship and may not be exhaustive.

\section{Review of the eigenstate thermalization hypothesis}\label{ETHreview}

%The foundation of ETH are revisited in detail before using these concepts in a quantum algorithm.

An initial system configuration is not necessarily in the equilibrium at the outset. The eigenstate thermalization hypothesis (ETH) is used here to justify thermalization in quantum systems. The key idea of ETH that is useful here is the ability for a system to remove the initial state dependence, so the formulation here will focus on that.  %But it should be expected to achieve this equilibrium average through some process, diffusive or otherwise. %The justification cited here is from random matrix theory but the results are expected to generalize to generic quantum systems.
%The many other works written on this topic do not generally formulate the problem in formal mathematical titles, and some papers may define some terms differently. The hope here is that this will help guide the future discussion by creating a list of terms that will be justified in the quantum circuit's case. Other definitions are possible.

%\subsection{Thermalization of a quantum systems}
%There is a remarkable fact of quantum physics related to the ability for a quantum system to thermalize. In classical physics, closed systems are expected to thermalize. 

%The quantum system does not necessarily accompish this. In the simplest treatment, the unitary nature of the time evolution allows for the unwinding of any state to its initial condition. ETH is a means of understanding how thermalization occurs in quantum systems.

Traditional quantum mechanics does not necessarily lead to a thermalized outcome. This will be reviewed first before adapting the argument to integrable and non-integrable systems. Random matrix theory is then used to derive the necessary equivalences behind Conjecture~\ref{ETH}.

\subsection{Introduction: Unitary dynamics of a quantum system}
We can identify that the time-evolution operator of an arbitrary input $\hat A$ on an arbitrary vector is
\begin{equation}\label{time_r}
e^{-i\hat At}|r\rangle=\sum_{p=1}^Nc_ke^{-iE_kt}|k\rangle=|\Psi(t)\rangle
\end{equation}
for some eigenbasis $\{|k\rangle\}$. When we go to take an expectation value of an operator $\hat{\mathcal{O}}$ with the time-evolved vector, 
\begin{equation}\label{timeavgintegrable}
\mathcal{O}(t)=\langle\Psi(t)|\hat{\mathcal{O}}|\Psi(t)\rangle=\sum_p d_{pp}|c_p|^2+\sum_{p\neq q}d_{pq}c^*_pc_qe^{i(E_p-E_q)t}
\end{equation}
with $\hat{\mathcal{O}}=\sum_{pq}d_{pq}|p\rangle\langle q|$. In the form of Eq.~\eqref{timeavgintegrable}, the starting-state dependence (captured by the coefficients $c_k$) is always be retained due to the unitarity of quantum physics. Choosing a different starting state with different coefficients $c_k$ will therefore be visible in any time-evolution. Further, the reverse time-evolution can always recover the original state. As stated, the quantum system does not thermalize.

The above behaviour will be applicable for integrable systems. 

\begin{definition}[Integrability]
An integrable problem is one where enough constants of motion are included for each conserved quantity, sufficient to solve the problem.
\end{definition}

Chaotic systems are generally only possible in non-integrable cases. This will be the case of interest here where the initial state can be lost in a time-evolution.%Quantum computing inherently seeks to solve problems that are non-integrable, since it is expected that only stabilizer circuits are efficiently simulated on the classical computer.% \cite{gottesmanknill}.

\begin{definition}[Non-integrable]
Any system that is not integrable.
\end{definition}

The puzzle that ETH is attempting to solve is to understand how the coefficients $c_k$ are forgotten during a time-evolution. If this cannot happen, then the equilibrium value of the system is not obtained even after a long time. ETH will not give a strict prescription for which systems thermalize. Instead, it is said that a system will thermalize if it satisfies ETH. In another sense, this distinction between integrable and non-integrable could be defined by which systems satisfy ETH and which do not.

%\subsection{Relationship to the trace}
%
%\begin{equation}\label{ETH}
%\overline{\langle\Psi(t)|\hat \Delta|\Psi(t)\rangle}\overset!=\frac1N\sum_{p=1}^N\langle p|\hat \Delta|p\rangle=\frac1N\mathrm{Tr}\left(\hat \Delta\right)
%\end{equation}
%where $p$ indexes eigenstates of $\hat A$.

\subsection{Random matrix theory justification for the eigenstate thermalization hypothesis}

ETH was properly formalized in Ref.~\cite{srednicki1994chaos,deutsch1991quantum}. The core statement is contained in Conjecture~\ref{ETH}. There was evidence for the concept before that. %A trio of cases where matrix elements of an operator can be computed form the foundation for the main conjecture. 
%Each of random matrix theory, semi-classics, and a statement by von Neumann point to the same behaviour for quantum systems. Even though each assumes conditions on the quantum system, the end result is assumed even for non-trivial cases ({\it e.g.} many interacting many-body systems) and suggests ETH. 
A complete proof for ETH does not exist at the present time, but it is backed up by a multitude of numerical evidence.

%\subsubsection{Random matrix theory}

The core statement for ETH was determined as a result of an explicit computation of the matrix elements of an operator $\hat{\mathcal{O}}$ based on random matrix theory (RMT). RMT was originally introduced in mathematical statistics \cite{kravtsov2009random,guhr1998random,mehta2004random} and then adopted for energy levels in physical systems \cite{wigner1993characteristic1,wigner1993characteristic2,dyson1962statistical}. The core assertion is that the energy spectra of a quantum system with interactions and of a sufficient size appear random.

Under RMT, the form of an operator is well approximated by the following result.

\begin{lemma}[Operator expectation values in random matrix theory]\label{RMTconj}
The eigenstate thermalization hypothesis defines a relationship for the matrix elements of an operator $\hat{\mathcal{O}}$ as \cite{deutsch2018eigenstate,d2016quantum,srednicki1999approach}
\begin{equation}\label{RMToperator}
\mathcal{O}_{mn}=\mathcal{O}(\bar E)\delta_{mn}+e^{-S(\bar E)}g(\bar E,\omega)R_{mn}
\end{equation}
where $\bar E=(E_m+E_n)/2$, $\omega=(E_m-E_n)/2$, and $R_{mn}$ is a random number. The correction to the first term contains a factor $\exp(-S(\bar E))$ which contains the entropy $S$ for the average energy $\bar E$. The entropy decreases quickly with system size in most cases, so the random numbers expressed by $R_{mn}$ are suppressed quickly with system size ({\it i.e.}, fluctuations decrease with system size). The function $g$ is expected to be a smooth function and in general all functions are taken to be smooth continuations of the discrete terms.
\end{lemma}

%\textcolor{red}{HERE}

The proof\footnote{Sometimes is this is called srednicki's ansatz \cite{d2016quantum}.} for this is based on RMT \cite{d2016quantum}. For brevity, the elements of RMT are introduced incrementally here in the context of other discussion and will not be re-derived here \cite{srednicki1994chaos,srednicki1999approach,berry1977regular,pechukas1983distribution,chirikov1985example,feingold1986distribution,wilkinson1987semiclassical,prosen1994statistical,eckhardt1995approach,eckhardt1995semiclassical,hortikar1998random}.% \cite{mehta}. %RMT supposes that for a window of eigenstates that the .... applies for few-body systems. % that do not include interactions

The relationship between $\mathcal{O}(\bar E)$ and the relevant expectation values in Conjecture~\ref{ETH} must now be established. Returning back to the original motivation of the study, the time averaged expectation value can be evaluated with the expectation of ergodicity in the classical limit \cite{reif2009fundamentals} since the entropy is maximized \cite{jaynes1957information,ash1965information}. The result is
%
%Combining the above argument with the form assumed in Eq.~\eqref{RMToperator} implies that an expectation value\footnote{The form chosen here is to match Liouville's theorem \cite{reif2009fundamentals}.}
\begin{equation}\label{quantumLiouville}
\overline{\mathcal{O}(t)}=\underset{t\rightarrow \infty}\lim\frac1t\int^t_0dt'\mathcal{O}(t')=\sum_m|c_m|^2\mathcal{O}_{mm}%=\mathrm{Tr}\left(\hat\Gamma_\mathrm{DE}\hat{\mathcal{O}}\right)
\end{equation}
where as $t\rightarrow\infty$, the oscillating terms on the off-diagonal elements average to zero provided that a single symmetry sector is used.

The end-quantity on the right of Eq.~\eqref{quantumLiouville} is worthy of its own definition as a new ensemble.

\begin{definition}[Diagonal ensemble]

A diagonal ensemble (DE) is defined as 
\begin{equation}\label{DEeq}
\overline{\mathcal{O}(t)}=\sum_m|c_m|^2\mathcal{O}_{mm}=\mathrm{Tr}\left(\hat\Gamma_\mathrm{DE}\hat{\mathcal{O}}\right)
\end{equation}
where an element of the density matrix $\hat\Gamma_\mathrm{DE}(m,n)=|c_m|^2\delta_{mn}$. The diagonal ensemble contains the knowledge of the initial wavefunction.

\end{definition}

The time averaging in Eq.~\eqref{DEeq} includes many states in the time-evolution, hence why the diagonal ensemble is an ensemble from Def.~\ref{ensembledef}, but it is still reliant on the coefficients of the initial state. It should (and does) appear unnatural to retain a relationship with the initial wavefunction based on the knowledge of thermalization in the classical limit, and it is only in the context of RMT can it be demonstrated concretely how that initial state dependence goes away. This hints that there should be a relationship that removes the initial state dependence.

Left in the form of Eq.~\eqref{DEeq}, the initial state dependence remains with no clue of how to remove it for a quantum system to thermalize. This would be undesirable to our main goal since it would leave the record of the initial wavefunction in place during the average. This would not match classical physics.

In the following, an additional definition of the average is required.
\begin{definition}[Average energy]\label{avgEdef}
The average energy of the system (energy as a microcanonical ensemble expectation value) is given by
\begin{equation}
\langle E\rangle\equiv\frac1{N}\mathrm{Tr}(\hat\Gamma_\mathrm{MC}\hat{\mathcal{O}})=\frac1{\mathcal N}\sum_mE_m
\end{equation}
%or equivalently in terms of the energies $E_m$ from Lemma~\ref{RMTconj} as
%\begin{equation}
%\langle E\rangle=\frac1{\mathcal N}\sum_{m=1}^{\mathcal N}E_m
%\end{equation}
which is taken over $\mathcal{N}$ ($\equiv\mathrm{Tr}(\hat\Gamma_\mathrm{MC}$)) states in the window (indexed by $m$) from Def.~\ref{MCdef} containing $\langle E\rangle$ and sufficiently wide around the average energy.
\end{definition}

So far, a connection between the diagonal matrix elements $\mathcal{O}_{mm}$ and $\mathcal{O}(\bar E)$ has been determined under RMT based on Eq.~\eqref{RMToperator}. At this point, we have no general relationship between $\mathcal{O}_{mm}$ and $\overline{\mathcal{O}(t)}$. However, %in order to make the connection of the main statement to show Conjecture~\ref{ETH} under RMT, the individual matrix elements must be related to a single value. 
%
%There is one intermediate step before  relating to the micro-canonical ensemble value. The relation derived first is between $\mathcal{O}(\bar E)$ and the value of the operator at the average energy. 
contained in the statement of the ansatz in Lemma~\ref{RMTconj} is the continuous nature of the variables involved. There is a useful corollary that can be defined that is useful for eventually deriving Conjecture~\ref{ETH}, which requires a relationship between $\mathcal{O}_{mm}$, $\mathcal{O}(\bar E)$, and $\langle\mathcal{O}\rangle_\mathrm{MC}$.

\begin{corollary}[Series expansion of the diagonal matrix elements]\label{ETHcor1}
Consider then an assumption in Lemma~\ref{RMTconj} that the variables were continuous and therefore can be expanded in a series expansion about some point. In this case, the useful expansion of the matrix elements $\mathcal{O}_{mm}$ would be to relate this to the average energy over the entire system, $\langle E\rangle$. The resulting expansion is
\begin{equation}\label{Omm_expand}
\mathcal{O}_{mm}=\mathcal{O}(\langle E\rangle)+\left.(E_m-\langle E\rangle)\frac{d\mathcal{O}}{dE}\right|_{\langle E\rangle}+\left.\frac12(E_m-\langle E\rangle)^2\frac{d^2\mathcal{O}}{dE^2}\right|_{\langle E\rangle}+\ldots
\end{equation}
which is shown to second order. It is important to notice that each of the elements on the diagonal of Eq.~\eqref{RMToperator}, $\mathcal{O}_{mm}$, are taken to be distributed about the the same zeroth order term in the series expansion of all terms regardless of $m$.
\end{corollary}

Using Corollary~\ref{ETHcor1}, the diagonal ensemble can be shown to be equivalent to $\mathcal{O}(\langle E\rangle)$.

\begin{corollary}[Diagonal ensemble equivalence under random matrix theory]\label{DEequivRMT}
The diagonal ensemble expectation value is equal to
\begin{equation}
\overline{\mathcal{O}(t)}\approx\mathcal{O}(\langle E\rangle)
\end{equation}
under the assumptions of RMT.
\end{corollary}

%The proof is provided in this case.

\begin{proof}
The core result in Eq.~\eqref{Omm_expand} can be inserted back into Eq.~\eqref{DEeq} to find
\begin{equation}
\sum_m|c_m|^2\mathcal{O}_{mm}=
\sum_m|c_m|^2\left(\mathcal{O}(\langle E\rangle)+\left.(E_m-\langle E\rangle)\frac{d\mathcal{O}}{dE}\right|_{\langle E\rangle}
+\left.\frac12(E_m-\langle E\rangle)^2\frac{d^2\mathcal{O}}{dE^2}\right|_{\langle E\rangle}
+\ldots\right)
\end{equation}
and we can analyze the result term by term. The first term is straightforwardly
\begin{equation}
\sum_m|c_m|^2\mathcal{O}(\langle E\rangle)=\mathcal{O}(\langle E\rangle)\sum_m|c_m|^2=\mathcal{O}(\langle E\rangle)
\end{equation}
since $\mathcal{O}(\langle E\rangle)$ is independent of $m$ and $\sum_m|c_m|^2=1$ by normalization of the quantum system. This does provide the proof of the main statement, but the other two terms should be shown to be either zero or vanish in the thermodynamic limit.

The second term has a deviation of the form 
\begin{equation}
\delta E_m=E_m-\langle E\rangle
\end{equation} 
where in general deviations of this form in statistical physics decrease as $1/\sqrt N$ \cite{pathria32statistical} by the central limit theorem \cite{reif2009fundamentals}. This means that each of $E_m$ terms get closer to the average value $\langle E\rangle$ with increasing system size.

%and can be identified from the sum over $m$ as
%\begin{equation}
%\sum_m\delta E_m=\sum_m(E_m-\langle E\rangle)=0
%\end{equation}
%which follows from Def.~\ref{avgEdef}. Thus, this term is zero in Eq.~\eqref{Omm_expand}

The final term in Eq.~\eqref{Omm_expand} contains a factor of $(\delta E_m)^2$. The deviations, $\delta E_m=(E_m-\langle E\rangle)$, from the average value have a well-defined thermodynamic limit in that they become negligible. Consider the generic behaviour of a quantum system (or quantum statistical system) that the uncertainty in an expectation value is related to 
\begin{equation}
\delta E=\sqrt{\langle\Psi|\hat H^2|\Psi\rangle-\langle\Psi|\hat H|\Psi\rangle^2}
\end{equation}
where the uncertainty for the Hamiltonian is used here. In a statistical system, we should expect that the uncertainty in the expectation value becomes more and more like a delta function as the system approaches the thermodynamic limit \cite{reif2009fundamentals}. This implies that $\delta E\rightarrow0$ for large systems. Since it scales less than the size of the number of elements of the system $N$ in this limit, this is known as subextensivity. Distributions in statistical physics are strongly peaked Gaussian functions, which justifies that the deviations go to zero.

Since only the leading order remains to be non-zero, the system will fluctuate about a mean value of $\mathcal{O}(\langle E\rangle)$ such that the oscillations are naturally damped out as the system tends to the thermodynamic limit. Small systems will have a fluctuation about the order zero value.
%
%This term is expected to tend to zero in the thermodynamic limit 
%
\end{proof}

Given a relationship between $\overline{\mathcal{O}(t)}$ and $\mathcal{O}(\langle E\rangle)$, there still must be found the connection to the micro-canonical ensemble. This can be seen to connect to the main result of Corollary~\ref{DEequivRMT}. 

\begin{corollary}[Equivalence micro-canonical ensemble under random matrix theory]\label{MCequivC}
The micro-canonical ensemble expectation value is equal to the diagonal ensemble
\begin{equation}
\mathcal{O}(\langle E\rangle)\approx\mathcal{O}_\mathrm{MC}
\end{equation}
under the assumptions of RMT.
\end{corollary}

\begin{proof}
In this case, return to Def.~\ref{microcanonicalensemble} and apply the expansion from Eq.~\eqref{Omm_expand} to obtain
\begin{equation}
\langle\hat{\mathcal{O}}\rangle_\mathrm{MC}=\frac1{\mathcal{N}}\sum_{E<E_k<E+\Delta E}\left(\mathcal{O}(\langle E\rangle)+\left.(E_k-\langle E\rangle)\frac{d\mathcal{O}}{dE}\right|_{\langle E\rangle}
+\left.\frac12(E_k-\langle E\rangle)^2\frac{d^2\mathcal{O}}{dE^2}\right|_{\langle E\rangle}
+\ldots\right)
\end{equation}
where $\mathcal{O}_{mm}$ from Eq.~\eqref{RMToperator} was re-labeled to match Eq.~\eqref{microcanonicalensemble} as $\mathcal{O}_{kk}=\langle k|\hat{\mathcal{O}}|k\rangle$.

The first term in the summation in the parenthesis is $\mathcal{N}$ copies of $\mathcal{O}(\langle E\rangle)$. The second term and third term vanish for the same reasons given in the proof for Corollary.~\ref{DEequivRMT}.
\end{proof}

Thus, to leading order
\begin{equation}
\mathcal{O}(\langle E\rangle)\approx\overline{\mathcal{O}(t)}\approx\mathcal{O}_\mathrm{MC}
\end{equation}
under the assumptions of RMT. The last two terms are the most important since they imply that core statement of Conjecture~\ref{ETH} is demonstrated under RMT. The hypothesis is supposed to hold for systems beyond matrices expressing randomness in their eigenvalue spectra, although that is not proven.

Having now identified the diagonal elements of a time evolution, the next natural question is to determine how the fluctuations of the variable defined by
\begin{equation}\label{sigmasquared}
\sigma^2=\underset{T\rightarrow\infty}\lim\frac1T\int_0^Tdt\Big(\mathcal{O}(t)^2-(\overline{\mathcal{O}})^2\Big)
\end{equation}
are at each individual time-step. Showing that the value remains close to the average value given above would imply that at each time-step the average value holds on average. This can be directly demonstrated with RMT as in the following.

\begin{corollary}[Sufficient condition for thermalization]
Under random matrix theory, a system satisfying the eigenstate thermalization hypothesis is a sufficient condition to explain thermalization.
\end{corollary}

\begin{proof}
The goal is to show that Eq.~\eqref{sigmasquared} is small for all times and as the system gets larger. This would imply that the fluctuations of the theory are small enough to indicate that a system satisfying Lemma~\ref{RMTconj}. 

First, expand the terms $\mathcal{O}(t)$ as
\begin{equation}
\sigma^2=\underset{T\rightarrow\infty}\lim\frac1T\int_0^Tdt\Big(\mathcal{O}_{mn}\mathcal{O}_{pq}c_m^*c_nc_p^*c_qe^{i(E_m-E_n+E_p-E_q)t}-\overline{\mathcal{O}}^2\Big)
\end{equation}
and expand using Eq.~\eqref{RMToperator}. The diagonal elements in the first term and $\overline{\mathcal{O}}$($=\mathcal{O}_\mathrm{MC}$ by ETH) cancel leaving only \cite{d2016quantum}
\begin{equation}\label{offdiagonal}
\sigma^2=\sum_{m,n\neq m}|c_m|^2|c_n|^2|\mathcal{O}_{mn}|^2
\end{equation}
where a time average is taken, removing the complex exponential term as in Eq.~\eqref{quantumLiouville}.

Applying RMT from Lemma~\ref{RMTconj} demonstrates that the maximum of the off-diagonal terms is the upper-bound of Eq.~\eqref{offdiagonal}. Since the off-diagonal terms under RMT scale as $\exp(-S(\bar E))$, then the summation vanishes in the thermodynamic limit since the entropy is expected to be large in that case for states in the bulk. This would demonstrate that the time-average expectation value is close to the thermalized value for large systems.% especially where the entropy $S$ is large.
\end{proof}

Thus, satisfying ETH demonstrates that the time-expectation value is close to the micro-canonical ensemble value. Thus, ETH under RMT directly demonstrates that the system is a sufficient condition to thermalize.

\subsection{Why is the micro-canonical ensemble average relevant here?}

Thermalization was defined at the outset to remove the dependence of the initial wavefunction from the final average in non-integrable systems. The equivalence to the micro-canonical ensemble is somewhat of a surprise in that the ensemble is often abstractly defined. Because of the equivalence with the canonical ensemble, one question is why report the results with the micro-canonical ensemble?

Of the quantities derived for Corollaries~\ref{DEequivRMT} and \ref{MCequivC}, only $\overline{\mathcal{O}(t)}$ and $\mathcal{O}_\mathrm{MC}$ are defined in the statistical physics literature. The definitions of $\mathcal{O}(\langle E\rangle)$ and the diagonal ensemble are defined specifically for the purposes of proof of the corollaries. So, in one sense leaving only the results of Corollary~\ref{DEequivRMT} stand on its own equates $\mathcal{O}(\langle E\rangle)$ with a specifically defined diagonal ensemble. The result is not general. Equating $\mathcal{O}(\langle E\rangle)$ to the micro-canonical ensemble does connect to a well-studied ensemble. Lacking a direct connection to the canonical ensemble, the proof leading to the micro-canonical ensemble is the preferred connection.

Other foundations not involving RMT can lead to a micro-canonical ensemble (see Ref.~\cite{deutsch2018eigenstate}), but ETH defined on RMT one of the most widely invoked. In one particular case, an argument in semi-classics \cite{berry1981quantizing,jarzynski1997berry} demonstrates that for a given energy that an ensemble of wavefunctions can be chosen with a Gaussian probability. Relatedly in this conjecture from semi-classics, an eigenstate looks as it chosen randomly from the ensemble. The resulting probability distribution of a given energy has an explicit form in this conjecture. The distribution is proportional to $\delta(E-\hat{\mathcal{H}})$ (for a Hamiltonian $\hat{\mathcal{H}}$) \cite{voros1976semi,jarzynski1997berry} which resembles a micro-canonical ensemble being defined over a narrow energy window ($E$). This was cited in Ref.~\cite{srednicki1994chaos} as demonstrating that micro-canonical ensembles can come from other techniques other than RMT and could be expected to show up as a general physical feature.

\section{Thermalization of quantum circuits}\label{ETHcircuits}

Broadly, circuits are composed of quantum operators and ETH was formulated on expectation values of generic operators. There is reason at the outset to expect ETH to extend to quantum circuits. %Enforcing full ergodicity in the final thermalized state would then give access to an equal superposition of eigenstates, from which a range of linear algebra problems can be computed.% Taking the small valued elements of the input operator to be zero, we can recover the micro-canonical average from the full trace.
The application of ETH to quantum circuits is extended in the following. 

%It is tempting to call the effect here a thermal quantum circuit proposition, but there is very little different from ETH here. 

%\begin{theorem}[Thermalization of a quantum circuit]
%\label{TQC}
%Averaging the time evolution of a circuit when using a quantum phase estimation and measuring over the identity operator produces a random draw of Gaussian states according to a statistical ensemble. This is the same justification as for the eigenstate thermalization hypothesis.
%\end{theorem}

%\begin{proof}
Consider a quantum circuit on two registers such that a time evolution operator $\exp(-i\hat At)$ is applied onto an arbitrary starting state. The time-evolved wavefunction of the form
\begin{align}
|\Psi(t)\rangle|E\rangle&=\mathrm{QPE}(e^{-i\hat At}\otimes\mathbb{I})|r\rangle|0\rangle^{\otimes m}\\
&=\sum_kc_k(e^{-i E_kt}\otimes\mathbb{I})|k\rangle|E_k\rangle\label{TEstate}
\end{align}
where $|r\rangle=\sum_kc_k|k\rangle$ in terms of eigenstates indexed by $k$. The auxiliary register containing $|E_k\rangle$ is discovered by the quantum algorithm when applying the QPE.

%The idea is to thermalize the circuit. Start with an super-operator
%\begin{equation}
%\check{\mathcal{O}}=\hat\Delta\otimes\hat\Upsilon
%\end{equation}
%which applies onto an eigenstate and the register containing its eigenvalue as%\footnote{}
%\begin{equation}\label{checkM_elements}
%(\langle p|\otimes\langle E_p|)\check{\mathcal{O}}(|q\rangle\otimes|E_q\rangle)=\Delta_{pq}f(E_p)%\delta(E_p-E_q)
%\end{equation}
%where the matrix elements $\Delta_{pq}=\langle p|\hat\Delta|q\rangle$ and $\hat\Upsilon$ acts as in Eq.~\eqref{Upsilon_operation}.\footnote{Based on how Eq.~\eqref{checkM_elements} is written, it may be tempting to add a term $\delta(E_p-E_q)$, but the ket and the dual wavefunction have a QPE and QPE$^\dagger$ term in them which is best viewed as only taking the state of $\hat\Upsilon$ and weighting each eigenstate. The eventual expectation value is taken over $|0\rangle$ states.} The output of the QPE reports a bitstring that requires a rule to turn into the eigenvalue. In this case, the $\hat\Upsilon$ operator implies that rule and turns the discrete bitstrings into a continuous variable. It is then supposed that the functions of that variable are smooth.

%We retain the delta function containing the energies to account for degeneracies that can appear in the eigenvalue spectrum. These will 

The operator $\hat A$ can often be thought of as having two separate pieces in anticipation of the use of two-registers on the quantum computer. One part ($\hat\Upsilon$, acting on an $m$-qubit register) is related to a function of the eigenvalues and $\hat\Delta$ is related to all other operators composing $\hat A$ ({\it i.e.}, Eq.~\eqref{Oexpect} would see $\hat\Delta=\hat\rho$, acting on an $n$-qubit register). The next step is to identify the time-dependent expectation value of the super-operator
\begin{equation}
\check{\mathcal{O}}=\hat\Delta\otimes\hat\Upsilon
\end{equation}
where $\check{\mathcal{O}}$ is the promotion of $\hat{\mathcal{O}}$ to be written explicitly on two registers as a composition of two matrices. The operator $\hat\Upsilon$ (defined on the $m$-qubit precision register) was defined in Def.~\ref{Upsilon_operation} and is a diagonal operator that weights each eigenstate by some factor $f(E_k)$.

The time evolution in this case is given explicitly by
\begin{equation}\label{timedep_expval_superM}
%\overline{
\langle\Psi(t)|\check{\mathcal{O}}|\Psi(t)\rangle
%}
=\sum_p|c_p|^2\Delta_{pp}f(E_p)+\sum_{p\neq q}c_q^*c_p\Delta_{pq}f(E_p)e^{i(E_p-E_q)t}%\delta(E_p-E_q)
\end{equation} 
where Eq.~\eqref{TEstate} was used. The thermalization of Eq.~\eqref{timedep_expval_superM} is accomplished through the average of the time-evolution, just as before.

The next step is to invoke ETH. We suppose that each of the terms dependent on $p$ are part of a distribution around an average energy $\langle E\rangle$. The ETH ansatz then indicates that each term dependent on $k$ loses its $k$-depdenence, and each becomes a uniform value, $\langle\check{\mathcal{O}}\rangle_\mathrm{MC}$. After all, there was no specification on the operator in ETH, so it should apply equally on $\check{\mathcal{O}}$. The time-average under ETH becomes
\begin{equation}\label{circuitETH}
\overline{\langle\Psi(t)|\check{\mathcal{O}}|\Psi(t)\rangle}
=\sum_p|c_p|^2\Delta_{pp}f(E_p)=\langle\check{\mathcal{O}}\rangle_\mathrm{MC}\sum_p|c_p|^2=\langle\check{\mathcal{O}}\rangle_\mathrm{MC}
\end{equation} 
where the remaining squared coefficients $|c_p|^2$ sum to 1 to preserve normality of the wavefunction. Note that the coefficients would only have summed to 1 independently upon measurement if the system register (size $n$ qubits) is disentangled from the precision register (size $m$ qubits), which is not the general case.%This is called the {\it diagonal ensemble}. 

The Eq.~\eqref{circuitETH} is presented assuming that ETH applies for the quantum circuit. In general, it would be required that the system is non-integrable.\footnote{The non-integrability may come in situations where the QPE is applied. It is not generally expected that knowledge of the first $m$ bits gives knowledge of the $m+1$th bit. Thus, the QPE can give non-integrability in a quantum circuit. More general, the Gottesman-Knill theorem \cite{gottesman1998heisenberg} indicates that not all circuits are classically simulable (only stabilizers are), which would be another source of non-integrability in a quantum circuit.} Because $\check{\mathcal{O}}$ is an operator and ETH was broadly derived for any quantum operator, the results from Sec.~\ref{ETHreview} immediately apply, keeping the statement short. The justification for ETH with RMT will be taken to be sufficient here, but it does establish that a quantum circuit can thermalize in the same way as ETH.

\subsection{Equal superposition of eigenstates}

The result in Eq.~\eqref{circuitETH} is formally the micro-canonical ensemble from ETH, which sums over a plurality of states. What is required for the quantum algorithm to be applicable in the general case is for the equivalent temperature of the canonical ensemble to be infinite ($\beta=0$). 

As was observed for the case of ETH in Sec.~\ref{ETHreview}, if one time-evolves a state with a perturbation on the operator $\hat A$, then there is no guarantee that a sum over all eigenstates is achieved, only that the micro-canonical ensemble is which is a plurality of states. 

If accomplished, this would make all states equally probable in the ensemble that the system is thermalized to. The equivalent wavefunction--that no longer needs to be prepared--is an equal superposition of eigenstates.  Another way to write this is to equate the trace of the input operator with the micro-canonical ensemble. This would replace $\overline{\langle \check{\mathcal{O}}\rangle}\approx\mathrm{Tr}\left(\check{\mathcal{O}}\right)$ instead of $\langle\check{\mathcal{O}}\rangle_\mathrm{MC}$ (or $\langle\check{\mathcal{O}}\rangle_\mathrm{MC}$ with $\beta\approx0$ in the canonical ensemble). 

\subsection{Generation of full ergodicity}

A natural question is how to modify $\hat A$ to kick the system hard enough to generate a fully ergodic thermalization. 
%
%However, the time-evolution is not constrained to be only with respect to $\hat A$. 
it is common practice in studies of quantum chaos to introduce stronger and stronger perturbations to turn an arbitrary model into a chaotic one. In that context, there are several modifications of the time evolved operator $\exp(-i\hat At)$ that can introduce full ergodicity. 

Modifications of $\hat A$ may be necessary for the general case here. If, for a fermionic example (operators $\hat c_{i\sigma}$ on site $i$ with spin $\sigma$), the operator has a symmetry in particle number and spin, it may be required to thermalize over an operator like $\hat A+\left(\hat c^\dagger_{i\sigma}\hat c^\dagger_{j\sigma'}+\mathrm{h.c.}\right)$ or potentially adding a term like $\hat c^\dagger_{i\sigma}\hat c_{j\sigma'}$. The exact schedule of when to apply the modified operator and when to apply $\hat A$ is not a strict prescription and may depend on $\hat A$.

For the type of problem that is being investigated here, note that many problems in linear algebra do not necessarily have a symmetry expressed. In this case, no modification may be required at all. Only a thermalization of $\hat A$ is required. The QPE will always be taken in the basis of the eigenstates, so this time-evolution is only useful for thermalization, not anything to do with the eventual weights of the expectation value.
%
%No matter the operator used to time-evolve, it is worthwhile to note that the QPE will be naturally described by a decomposition in the eigenstates of $\hat A$ since the operator used in the QPE is always $\exp(i\hat A)$.
%
%
%

With regards to the precision of the time evolution operator, this is an instance where numerical imprecision can potentially aid the thermalization of the quantum circuit. Imprecise matrix elements (or errors on the quantum computer) can create non-zero elements and cause transitions between symmetry sectors.  So, the quantum computation does not necessarily need to be noise-free, making this algorithm not necessarily only for the fault-tolerant era.\footnote{The resource requirements for the QPE would still need to be overcome.} %However, the writing of the operator $\exp(i\hat A)$ will require sufficient connectivity for the necessary decomposition of the operator. 

\subsection{Recovery of the micro-canonical ensemble}
The micro-canonical ensemble was defined over a window of energy states. If the goal is to thermalize to a state that has equal probability over all eigenstates in all symmetry sectors, then one can ask whether the micro-canonical expectation value is well defined according to Def.~\ref{MCdef}. 

As written, the fully ergodic state that the system will thermalize to ({\it i.e.}, $\mathrm{Tr}(\check{\mathcal{O}})$ over all eigenstates) is not a micro-canonical ensemble since all states must be summed instead of a plurality of states. %This raises a natural question: why rely on the eigenstate thermalization hypothesis if the micro-canonical ensemble is not recovered? Indeed, all of the evidence that we have for ETH relies on the generation the micro-canonical ensemble in the time-average. 

However, all of these computations here are phrased in terms of the QPE which imposes a finite precision on the problem. The micro-canonical ensemble is recovered during the QPE step for all practical purposes.
%One can argue an additional point not represented so far in the discussion of ETH will recover the mico-canonical ensemble. In a computation, a finite precision will be chosen and values in the problem below this precision can be taken to be zero. 
One could argue that at finite precision, there is still a definition of a micro-canonical ensemble. The window defined for this micro-canonical ensemble is best phrased in terms of the precision of the problem, but this is not different from an energy window based on the following.

%. The overarching goal is to compute Eq.~\eqref{conjecture1}.

%As phrased for a purely physical system, it is true that the micro-canonical ensemble is lost, but this is not true in the computational sense. 
Granularly, there are values in the computation much larger than a precision $\varepsilon$ that is chosen and numbers that are far smaller. For simplicity, the values that are approximately $\varepsilon$ are not included which is assumed to still approximate the full value. This assumption likely is most directly applicable when the precision is chosen to be smaller than the largest values in the system.

%\textcolor{red}{Requirements of the micro-canonical ensemble based on the definition with continuity}

The density matrix for the computation, $\hat\Gamma$, would then be defined as
\begin{equation}\label{Gamma_precision}
\Gamma_{kk}\approx\begin{cases}
1 & f(E_k)\langle k|\hat\rho|k\rangle>\varepsilon\\
0 & \mathrm{otherwise}
\end{cases}
\end{equation}
which introduces a precision into Eq.~\eqref{dualform}. The finite value of the precision creates the window over which the micro-canonical ensemble is summed over in cases where finite precision is requested, which is the practical case on the quantum computer. Hence, the QPE recovers the window that is characteristic of the micro-canonical ensemble that was lost when performing the full ergodic thermalization in the time-evolution operation.

The position taken in this work is that the final result is still a micro-canonical ensemble (defined by the finite precision of the problem) and that the time-evolved expectation value still satisfies ETH.

%An approximation is given in Eq.~\eqref{Gamma_precision} because of cases where some of the value $f(E_k)\langle k|\hat\rho|k\rangle$ is truncated by the precision, but the normal expectation for those cases in an accurate calculation is that they are small in magnitude anyway.

\section{Equal superpositions of eigenstates as a resource from thermalization}%} in a quantum circuit}

The above concepts of thermalization are used to create--on average--an equal superposition of eigenstates. The QPE can then be used to obtain a weighted trace of an initial input operator.

These two algorithms will be referred to as ETH-$\Sigma$ following the naming convention in Ref.~\cite{QGLD}. There are two forms that can be constructed based on this idea. One is a form based on an input vector, which would only be efficient when the operations to construct it are $O(n)$. The other form, the operator form, recasts all wavefunction preparation as density matrix operators and therefore avoids lengthy preparation on the quantum computer.

\subsection{Vector form}

An algorithm is now developed to find the expectation value of the inverse of an operator. What is needed is an arbitrary starting wavefunction $|r\rangle$ that is time evolved, as in Eq.~\eqref{time_r}. %From ETH, we can use Eq.~\eqref{ETH} to convert the time-average of this evolved wavefunction into the sum or trace over all eigenstates. 

The starting point is to create the arbitrary, non-eigenstate $|r\rangle$, and it is assumed that $|r\rangle$ is efficiently available on the quantum computer by some technique  ({\it e.g.}, Refs.~\cite{baker2021lanczos,bakerPRA24}) in order to fully implement the following.% (see Algorithm~\ref{ETH_vector}). 
%The starting state is then time evolved as in 

The generation of the term $1/E_k$ can be taken to be given by a quantum phase estimation (QPE) and another diagonal operator $\hat \Upsilon$ which contains the inverses of all eigenvalues. Note that the QPE discovers all eigenvalues, not to just one as in the standard use case \cite{kitaev1995quantum,nielsen2010quantum}.

The state that must be generated is
\begin{equation}
|\psi\rangle=(\mathrm{QPE}^\dagger)\left(I\otimes\sqrt{\hat\Upsilon}\right)(\mathrm{QPE})|\Psi(t)\rangle|0\rangle^{\otimes m}
\end{equation}
where $|0\rangle^{\otimes m}$ is an auxiliary register for the QPE to store the eigenvalue, $I$ is the identity, and $\sqrt{\hat\Upsilon}$ acts as
\begin{equation}\label{sqrtTau}
\sum_kc_k(t)\left(I\otimes\sqrt{\hat\Upsilon}\right)|k\rangle|E_k\rangle=\sum_k\frac{c_k(t)}{\sqrt{E_k}}|k\rangle|E_k\rangle
\end{equation}
where $c_k(t)=c_k\exp(-iE_kt)$, and the inverses of the eigenvalues are obtained from the QPE. The entries can potentially include negative eigenvalues which could be shifted depending on how $\hat A$ was implemented. 

A swap test with the input vector $\langle\Phi|$ will generate the overlap squared between the two vectors, $|\langle\Phi|\psi\rangle|^2$, and scales as $O(\varepsilon^{-2})$ \cite{barenco1997stabilization,buhrman2001quantum}. The time-averaged quantity will generate the sum over all eigenstates from Eq.~\eqref{ETH}.

The total complexity cost to arrive at this solution is $O(T/\varepsilon^2)$ where $T$ is the complexity to implement a decomposed time-evolution. The cost $T$ of the time evolution is expected to be $O(\tau\log_2N)$ for some thermalization time $\tau$ gates and $\log_2N$ operations generally \cite{low2023complexity}. 

The QPE here appears as a sub-leading cost $O(\varepsilon^{-1})$ \cite{kitaev1995quantum,nielsen2010quantum}. 
Another factor $M\equiv2^m$ assumes the worst-case scaling for a diagonal operator $\hat\Upsilon$ \cite{bullock2008asymptotically,welch2014efficient} but is a sub-leading exponential in terms of the overall problem size $N\ggg M$. 

The operator $\hat\Upsilon$, as with the quantum singular value decomposition \cite{gilyen2019quantum}, requires some methods to implement it. %But it should be noted that not all digits of precision may be required for all problems. Since we relate the precision over $m$ qubits to the precision as $\varepsilon=2^{-m}$, we accumulate another factor of $1/\varepsilon$. 

The algorithm should be expected to be exponentially costly in terms of qubits representing the precision requested in the algorithm, but often this number $M$ is far less than the total Hilbert space size $N$. The dominant source of error in the algorithm is the QPE as will be discussed after surveying a second way to use fully ergodic thermalization as a resource in an operator form.

%this is not dependent on condition number as explained in the discussion around Fig.~\ref{QPIE}.

\subsection{Operator form}

The previous vector-form of the operator assumes that we have the state $\Phi$ as a vector on the quantum computer. If instead this is captured by an operator $\hat\Delta=|\Phi\rangle\langle\Phi|$ then a similar procedure can be defined to take an expectation value of this operator. 

The generalization of the vector formulation is to then take the time-averaged expectation value of the form
\begin{align}
\langle B\rangle&=\overline{\langle\Psi(t)|\langle0|^{\otimes m}(\mathrm{QPE}^\dagger)\left(\hat\Delta\otimes\hat\Upsilon\right)(\mathrm{QPE})|\Psi(t)\rangle|0\rangle^{\otimes m}}\nonumber\\
&=\overline{\sum_{pq}\langle q|\langle E_q|c_q^*(t)\left(\hat\Delta\otimes\hat\Upsilon\right)c_p(t)|p\rangle|E_p\rangle}
\label{opform}
\end{align}
where $\hat\Upsilon$ produces a weight of $f(E_k)=1/E_k$ for the inverse instead of the square root as before. This is a time-average of the expectation value of the operator $\left(\hat\Delta\otimes\hat\Upsilon\right)$ over $\Psi(t)$ with a QPE used to determine the eigenvalues of the superposition.

\subsection{Weights from a quantum phase estimation}

Consider the relationship between the input matrix $\hat A$--see Eq.~\eqref{operatorA}--and its inverse
\begin{equation}\label{operatorA}
\hat A^{-1}=\sum_k\frac1{E_k}|k\rangle\langle k|
\end{equation}
which would give $\hat A\hat A^{-1}=\hat A^{-1}\hat A=\mathbb{I}$. The following will create a framework for any possible function of the eigenvalue, $f(E_p)$, not just the inverse as presented here. 

%Returning to Eq.~\eqref{expY_sump}, this allow for an expectation value as $\langle B\rangle=\sum_k\langle k|\hat\Delta|k\rangle/E_k$.

The quantum computer has an operation that can discover the eigenvalues called the quantum phase estimation (QPE), which is defined in Def.~\ref{QPE}.

%In general, the QPE can be defined as follows.

The circuit by which this is accomplished is given in Fig.~\ref{QPEcircuit} \cite{nielsen2010quantum}. A register of $m$ qubits (shown, $m=3$ in Fig.~\ref{QPEcircuit}) have a Hadamard gate applied
\begin{equation}
\hat H=\frac1{\sqrt2}\left(\begin{array}{cc}
1 & 1\\
1 & -1
\end{array}\right)
\end{equation}
and then a gate $\hat U=e^{i\hat A}$ is applied to generate a series approximation for the bitwise representation of the eigenvalue. After the inverse quantum Fourier transform (QFT), The phases on each of the $|1\rangle$ states are represented on qubits $|E_p\rangle$. This is represented as a bitstring, but there is in principle a map between the bitstring and the eigenvalue. %If represented on the interval [0,1], then the map is a positive operator value measurment (POVM). 

%The number of qubits of precision correspond to digits in a bitwise representation of the eigenvalue shifted and scaled to the interval $[0,1]$ as a positive operator value measurment (POVM). 

%There are two extensions that are worth noting. One is that the input eigenstate can be approximated. In this case, the distribution of probabilities around the derived eigenvalue, $E_p$, is a distribution about that value. The QPE can also be applied in superposition of the form Eq.~\eqref{superposition}.
%
%The core assertion that allows for the construction of Eq.~\eqref{} is that the thermalization of an arbitrary input state $|r\rangle$ and use the QPE 

The weighting matrix can then be measured as
\begin{equation}
(\langle\Xi|\otimes\langle E|)(\mathbb{I}\otimes\hat\Upsilon)(|\Xi\rangle\otimes|E\rangle)
\end{equation}
where $|E\rangle$ is an $m$-qubit register of eigenvalues $E_p$ in superposition ({\it i.e.}, sum of Eq.~\eqref{QPEaction}). This form effectively takes the expectation value of any function of the input matrix $f(\hat A)$. The equal superposition of eigenstates is represented here as a single pure state for ease of presentation. This would represent the algorithm for a single preparation of $|\Xi\rangle$ instead of the time-average as is proposed here to access it.

%The algorithm steps are contained in Algorithm~\ref{}.
%The main goal of this paper is therefore contained in the following proposition.
%
%\begin{proposition}[Measurements of functions of operators]
%The measurement of an operator, $f(\hat A)$--or a function of that operator, $f(\hat A)$--can be accomplished in $m$-qubits with the appropriate selection of $\hat \Upsilon$ and an equal superposition of eigenstates.
%\end{proposition}

Doing so would establish an algorithm that is poly-logarithmic, provided that the input operator is well-represented for the quantum algorithm, and complete the sense of solving many problems of the form Eq.~\eqref{Oexpect} (hence, a BQP algorithm is sought here).\footnote{The notion of a BQP-complete algorithm is available for classes of linear algebra here and can be demonstrated if there is no many-body localized behaviour in the thermalization of the circuit.}

The implementation of $\hat \Upsilon$ is not the focus here. Diagonal matrices are often easy to implement on quantum devices. 

\section{Examples}\label{examples}

Several examples are given for potential applications of the algorithm in the field of linear algebra. All of these are formulated in the operator form since the application of $\hat\Delta$ and $\hat\Upsilon$ would require significant implementation steps in the vector form. The operator form merely measures over them, which can be accomplished in the near-term.

%As a first example, only the thermalization of an operator $\hat\Delta$ is considered, with the form of the time evolution to be $\hat A\neq\hat\Delta$. This is not a firm requirement since we expect the QPE that we often want to apply to not be analytic.

\subsection{Thermalization of a small operator}

Take $\hat A=\sigma^z$, the Pauli-matrix \cite{townsend2000modern}. A random state $|r\rangle$ is time-evolved according to $\exp(-i\sigma^z\delta t)$ with time-step $\delta t$. Importantly, each new time-evolution starts from the same random vector. In this case we choose an equal superposition of all states initialized by applying Hadamard gates, $H$, on each qubit, $H^{\otimes n}$. The expectation value of
\begin{equation}\label{sum_derivs}
\overline{\langle \Psi(t)|\hat\Delta|\Psi(t)\rangle}\approx\mathrm{Tr}(\hat\Delta)/N=1/\sqrt2
\end{equation}
which is the correct value for the system to thermalize to when $\Delta$ is properly normalized as ($\mathbb{I}$ being the idenity)
\begin{equation}
\hat\Delta=\frac1{\sqrt2}\left(\begin{array}{cc}
1 & 1\\
1 & 1
\end{array}\right)=\frac1{\sqrt2}(\mathbb{I}+\sigma^x)\equiv\mathcal{I}_{2\times2}
\end{equation}
which computes the sum of all gradients in a system. In a numerical test, this was achieved out to 6 digits for $10^5$ samples and verified in more than 100 starting initial states. The $\hat\Delta$ operator here can be prepared with the QFT for general circuits as in Fig.~\ref{QFT_Delta}, but Hadamard gates can be used in this small example. The QPE that is missing here effectively weights the diagonal elements of $\hat\Delta$ with some function related to $E_p$, which projects the problem back onto ETH. Computations were performed with the DMRjulia library \cite{bakerCJP21,dmrjulia1}.

There are several implementations of interest for this matrix that involve non-trivial operators $\hat \Upsilon$. These examples are in some cases trivial, but they illustrate the general ability for the algorithm to solve many problems.

\subsection{Measuring the trace of a matrix}

The generation of the equal superposition of eigenstates and using a form of $\hat \Upsilon$ with only the eigenvalues as
\begin{equation}
\hat\Upsilon=\mathrm{diag}(E_1,E_2,\ldots,E_M)
\end{equation}
would give $\langle \hat A\rangle$ upon measurement and averaging of the time-evolved results.% where $\mathbb{I}$ is the identity matrix.

\subsection{Measuring an inverse matrix}
Equally, using the inverse of the eigenvalues would give the inverse matrix,
\begin{equation}
\hat\Upsilon=\mathrm{diag}\left(\frac1{E_1} ,\frac1{E_2} ,\ldots,\frac1{E_M}\right)
\end{equation}
which can be represented as an operator applied on $m$ qubits.

\subsection{Measuring the logarithm determinant}
Finding the logarithm determinant of a matrix can be found by converting% \cite{}
\begin{equation}
\log\mathrm{det}(\hat A)=\mathrm{Tr}(\log\hat A)
\end{equation}
where we must use
\begin{equation}
\hat\Upsilon=\mathrm{diag}(\log E_1,\log E_2,\ldots,\log E_M)
\end{equation}
then the result is the sum of the logarithmic terms, which is equivalent to the logarithm-determinant.

\subsection{Determinant of a matrix}
The exponential of the previous result for the logarithm-determinant will work here.

\subsection{Gradient of the logarithm-determinant}
In Ref.~\cite{QGLD}, an efficient representation of the quantum gradient of a logarithm-determinant (QGLD) was shown to be poly-logarithmic with a reasonable scaling in terms of precision. However, the form of the operator was not readily available. Under the framework provided here, we can provide an efficient implementation of the operator required to obtain the gradients here.

The gradient of a logarithm-determinant has near ubiquitous application in physics. It formally relates elements of the matrix with that of the inverse as
\begin{equation}
y_{ij}=\frac{\partial}{\partial x_{ij}}\ln\mathrm{det}\left(\hat X\right)
\end{equation}
for some element of $\hat X$ denoted as $x_{ij}$ and the element of the inverse $\hat Y=\hat X^{-1}$ as $y_{ij}$. Much of thermodynamics follows its structure if we set $\mathrm{det}\left(\hat X\right)=Z$ where $Z$ is the partition function and where the elements of the inverse $y_{ij}$ relate to observable quantities.

Here, the QGLD is modified to be amenable to a single operator in the ETH-$\Sigma$ algorithm developed here.\footnote{Ref.~\cite{QGLD}'s results are formulated around the quantum gradient algorithm which proposes an explicit unitary formulation of the problem. The formulation here is only simpler for the operator form here.}
%In terms of the eigenvalues $E_k$, this can be written as
%\begin{equation}
%y_{ij}=\sum_{k=1}^N\frac{\delta E_k(x_{ij})}{E_k}
%\end{equation}
%where
%\begin{equation}
%\delta E_k(x_{ij})\equiv\frac{\partial E_k}{\partial x_{ij}}
%\end{equation}
%which is abbreviated for clarity.

If an input operator $\hat X$ has derivatives of each value of the matrix that are contained in a matrix $\hat \Delta$, then the derivative would be defined as
\begin{equation}
%\frac{\partial \hat X}{\partial x_{ij}}
\nabla_{x_{ij}}\hat X=\frac{\left(\hat X+\epsilon\hat \Delta\right)-\hat X}\epsilon=\hat\Delta
\end{equation}
following the fundamental theorem of calculus. This can be reinterpreted to mean that each element has its derivative taken. Using the same form of $\hat\Upsilon$ as for the inverse with the appropriate $\hat\Delta$ operator gives the gradient of the logarithm determinant \cite{QGLD}.

Of the several important results in Ref.~\cite{QGLD}, two can guide the implementation of $\hat \Delta$. In Ref.~\cite{QGLD}, a small parameter $L$ from the quantum gradient algorithm would reveal only the operator $\hat \Delta$ only in the limit as $L\rightarrow0$, instead of perturbing the elements of $\hat X$ themselves. The second result from Ref.~\cite{QGLD} is that the general form of the $\hat\Delta$ matrix is
\begin{equation}
\hat \Delta\equiv|\Phi\rangle\langle\Phi|
\end{equation}
for some general wavefunction $\Phi$, although this operator does not need to be symmetric in general ({\it i.e.}, it could be $|\Phi\rangle\langle\Psi|$ for some other function $\Psi$, but in this form, one would need to determine real and imaginary parts of two separate measurements of $\hat\Delta=|\Phi\rangle\langle\Psi|\pm i|\Psi\rangle\langle\Phi|$). Ref.~\cite{QGLD} also establishes that the $\hat\Delta$ operator can be computed best when an equal superposition of eigenstates is available.

With this new capability to represent $\hat\Delta$ on its own, we can take the simple case of
\begin{equation}
\hat\Delta=\mathcal{I}=\left(\begin{array}{cccc}
1 & 1 & \ldots & 1\\
1 & 1 & \ddots & 1\\
\vdots & \ddots & \ddots & \vdots\\
1 & 1 & \cdots & 1\\
\end{array}\right)=\left(\hat H|0\rangle\right)^{\otimes n}\left(\langle0|\hat H\right)^{\otimes n}
\end{equation}
as a uniform operator of size $N\times N$ and note that its representation in Fig.~\ref{QFT_Delta}. By taking a quantum Fourier transform (QFT) \cite{nielsen2010quantum}, one can propose an operator to be constructed to make $\mathcal{I}$.

The above discussion will apply to an operator form of the algorithm given here, which involves measuring over the operator $\hat\Delta$. To obtain the vector form, one must prepare the wavefunction $\Phi$ on the quantum computer, which can be accomplished for $\mathcal{I}$ with Hadamard gates.

\begin{figure}
\begin{center}
\begin{quantikz}[column sep=14, wire types = {q, q, q, q}]
\lstick{$\ket{0}$}&\qwbundle[Strike Height= 0.2]{}&\gate[4]{\mathrm{QFT}} &\phase{} &\gate[4]{\mathrm{QFT}^\dagger}& \\
\lstick{$\ket{0}$}&\qwbundle[Strike Height= 0.2]{}& &\phase{}  \wire[u][1]{q} & & \\
\lstick{$\ket{0}$}&\qwbundle[Strike Height= 0.2]{}& &\phase{} \wire[u][1]{q} & & \\
\lstick{$\ket{0}$}&\qwbundle[Strike Height= 0.2]{}& &\gate{\hat\phi} \wire[u][1]{q} &&
\end{quantikz}
\end{center}
\caption{Implementation of $\hat \Delta=\mathcal{I}$ (a matrix of all ones) from the quantum Fourier transform. The operator $\phi$ makes a value of 1 in a diagonal matrix $\hat \phi=\mathrm{diag}(1,0,0,0,\ldots)$ so that $(\mathrm{QFT})^\dagger\hat \phi(\mathrm{QFT})=\mathcal{I}$.
\label{QFT_Delta}
}
\end{figure}

\subsection{Discussion}

The breadth of topics that can be solved with ETH-$\Sigma$ covers several fundamental operations in linear algebra. This lends to the notion of completeness since it applies in this case. Future works will focus on other cases where the algorithm applies to establish how complete it is.

To the identification of algorithms in the bounded quantum-polynomial (BQP) complexity class, the only possibility for this algorithm to be not poly-logarithmic is when the thermalization time is not of the order of the number of qubits, as would be the case for systems displaying MBL. However, the identification of MBL in physical system is mainly for low-dimensional cases and is not expected to be the general case. 

In general, the expectation is that the algorithm can be applied on even integrable cases since the map between the quantum state and the eigenvalue through the QPE should not expected to be integrable in general.

\section{Sources of error and their mitigation}\label{QPEissue}

There are a few sources of error in ETH-$\Sigma$. %The first two discussed are not necessarily the dominant sources of error. Errors from the QPE may play a role, but likely can be guarded against in actual implementation in principle.
%
%\subsection{Errors in thermalization}
There is the possibility that the thermalization itself introduces error; however, we will assume that running the algorithm longer will alllow for these errors to be reduced. Some fluctuations in the time-average is expected, but on average the systems should approach the proper value in $O(\log_2N)$ excepting in the pathological cases mentioned earlier.

%\subsection{Less dependence on the condition number}

The condition number is an important quantity for classical computation and is defined as follows.
\begin{definition}[Condition number]
The condition number of a matrix is defined as the ratio between the highest and low eigenvalues of the matrix
\begin{equation}
\kappa=\frac{E_1}{E_M}=\|\hat A\|\cdot\|\hat A^{-1}\|
\end{equation}
where on the quantum computer we can take $E_1\approx1$ and $E_M\approx\varepsilon$ to give $\kappa=1/\varepsilon$.
\end{definition}

In the algorithm here, because the initial wavefunction can be input as an operator, the inherent errors on the quantum computer will not affect the density matrix of the initial system. The standard relationship between the error of the vector $\mathbf{x}$ and the input $\mathbf{b}$ is \cite{golub1996matrix}
\begin{equation}
\frac{\delta x}{\|\mathbf{x}\|}=\kappa\frac{\delta b}{\|\mathbf{b}\|}
\end{equation}
where $\delta x$ ($\delta b$) is the error in the vector $\mathbf{x}$ ($\mathbf{b}$), and $\| \mathbf{x}\|$ ($\| \mathbf{b}\|$) is the modulus in the vector $\mathbf{x}$ ($\mathbf{b}$). %The condition number $\kappa$ is defined as follows.

Since, in principle, $\mathbf{b}$ can be implemented without the inherent errors on the quantum computer ($\delta b=0$), this should not contribute to the error of the system since $\hat\rho$ is classically known. So, the normal error mechanisms for this dual algorithm presented here will be different than some previously considered methods.\footnote{Therefore, the choice is made to represent the scaling of the algorithm with the inverse precision $1/\varepsilon$ where this is replaced by the condition number in many other published works.}

%On the classical computer, the increase in precision corresponds in a decrease in error

%\subsection{Quantum phase estimation of the inverse eigenvalues}

When performing the QPE, the bit-string that is reported from the subroutine reports the first bits which are the largest ones. If we take a bit-string of the form $0.x_1x_2x_3\ldots$, then we obtain first $x_1$ followed by $x_2$ and so on. This implies that the largest bits are obtained first and that more operations on the QPE are required to obtain the smaller bits. If the condition number of the matrix is large, then the lowest eigenvalues for an inverse problem will be the most important. Thus, the QPE would require many applications of the operator $\exp(i\hat A)$ in order to find the lowest digits. A remedy for this issue will be discussed in future works but it is common to many algorithms relying on the QPE.

\section{Conclusion}

Generating an equal superposition of eigenstates on the quantum computer was proposed using thermalization. The eigenstate thermalization hypothesis relates the time-averaged expectation value to a micro-canonical ensemble. Applying these concepts in a quantum circuit means that time-evolving an operator that sufficiently kicks the system and using the quantum phase estimation can create a state that contains both contains (on average) an equal superposition of eigenstates and the eigenvalues of each eigenstate. A subsequent measurement gives access to general expectation values in quantum mechanics. The time to thermalize the circuit is not generally as long as the Hilbert space size, implying an efficient algorithm. 

%Thermalization as a concept in 

%Formally, the thermal quantum circuit proposition is established, but the current justification relies on the eigenstate thermalization hypothesis. So, they are synonymous here but future innovations on understanding the mechanism behind thermalization in quantum systems is rigorously proven or understood from another viewpoint, then the results here should still be applicable.

\section{Acknowledgements}

The author thanks Luca D.~Adams, Jeffrey Morais, Olivia Di Matteo, Dominic Largoza, Raphael C.~Kelly, Jaimie Greasley, Mirko Amico, Marek Korkusinski, Brooke Tvermoes, Barry Sanders, Irina Paci, Anne Najdzionek, Andrew MacRae, Karl Thibault, Jesse Mumford, Michael O.~Flynn, Negar Seif, Nishanth Baskaran, Arif Babul, and Doug Johnstone for useful conversations. Some figures were prepared with the Quantikz package \cite{kay2018tutorial}. The author continues to thank and remember David Poulin.

This research was undertaken, in part, thanks to funding from the Canada Research Chairs Program (CRC-2021-00257). This work has been supported in part by the Natural Sciences and Engineering Research Council of Canada (NSERC) under grants RGPIN-2023-05510 and DGECR-2023-00026.
% This section is a list of funder names and grant numbers

%\roles{Sample text inserted for demonstration.}
% List author names and the contributions made to the article, using terms from the NISO Contributor Roles Taxonomy (CRediT) https://credit.niso.org

%\data{Sample text inserted for demonstration.}
% For more information on IOP Publishing's research data policy see: https://publishingsupport.iopscience.iop.org/questions/research-data/

%\suppdata{Sample text inserted for demonstration.}

%\section*{References}

%\bibliographystyle{natbib}%apsrev4-1}
\bibliographystyle{unsrt}
\bibliography{QBLR,RQBLM,TEB_papers,TEB_books,lanczos_refs,refs}

\begin{appendix}

%\section{Evidence for the eigenstate thermalization hypothesis}
%
%Semiclassics
%
%Random matrix theory
%
%\section{Random matrix theory}
%
%\subsection{Predictions from random matrix theory}

\end{appendix}

\onecolumngrid

\begin{appendix}

\end{appendix}

\end{document}